\makeatletter
\providecommand\reserveinserts[1]{}
\makeatother
\documentclass[AMA,Times2COL]{WileyNJDv5} 

\articletype{Original article}

\received{Date Month Year}
\revised{Date Month Year}
\accepted{Date Month Year}
\journal{Journal}
\volume{00}
\copyyear{2023}
\startpage{1}

\usepackage{mdframed}

\sethlcolor{white}

\begin{document}

\title{Adaptive PD-like Fault Tolerant Control for a Class of Nonlinear Systems}

\author[1]{Alessio Dallabona}

\author[2]{Vicenç Puig}

\author[1]{Mogens Blanke}

\author[1]{Dimitrios Papageorgiou}

\authormark{DALLABONA \textsc{et al.}}
\titlemark{Adaptive PD-like Fault Tolerant Controller for a Class of Nonlinear Systems}

\address[1]{\orgdiv{Department of Electrical and Photonics Engineering, CREA Group}, \orgname{Technical University of Denmark}, \orgaddress{\state{Kgs. Lyngby}, \country{Denmark}}}

\address[2]{\orgdiv{Department of Automatic Control (ESAII)}, 
\orgname{Universitat Politècnica de Catalunya (UPC)}, 
\orgaddress{\city{Barcelona}, \country{Spain}}}

\corres{Corresponding author Alessio Dallabona, \email{aldall@dtu.dk}}


\abstract[Abstract]{The article presents a nonlinear adaptive control strategy for a class of feedback-linearizable systems subject to process faults and disturbance. The controller is derived through Lyapunov analysis and incorporates an adaptive proportional gain within a PD-like structure, preserving simplicity while providing nonlinear adaptive behavior. A residual-based gating mechanism activates the adaptation only when the tracking error exceeds a prescribed threshold, preventing unnecessary gain updates during nominal operation and improving robustness to noise. The proposed design yields uniform global ultimate boundedness of the closed-loop signals, with explicit expressions for the steady-state tracking bound. Practical tuning guidelines for the adaptive parameters and Lyapunov matrix selection are provided to ensure consistent gating behavior. Simulations on a precision positioning system illustrate effective compensation of process faults, rapid gain escalation during degraded conditions, and smooth reversion to nominal performance, demonstrating that the method enhances fault tolerance while retaining the structural simplicity of PD control.}


\jnlcitation{\cname{%
\author{Dallabona A.},
\author{Puig V.},
\author{Blanke M.}, and
\author{Papageorgiou D.}}.
\ctitle{Adaptive PD-like Fault Tolerant Controller for a Class of Nonlinear Systems.} \cjournal{\it Int J Robust Nonlinear Control.} \cvol{2025;00(00):1--18}.}

\maketitle

\renewcommand\thefootnote{}
\footnotetext{\textbf{Abbreviations:} FTC, fault tolerant controller; FL, feedback linearization; NL, nonlinear.}

\renewcommand\thefootnote{\fnsymbol{footnote}}
\setcounter{footnote}{1}

\section{Introduction}
\label{sec:intro}

Control architectures grounded in  PID-like structures remain central to industrial practice because they combine simplicity, robustness, and minimal model dependency~\cite{alexandrovAdaptivePIDControllers2014}. 
Despite the availability of sophisticated nonlinear and adaptive control techniques, PID-based solutions perform reliably across a wide spectrum of nonlinear and uncertain systems, provided that suitable tuning or augmentation mechanisms are employed. 
This persistent effectiveness motivates the ongoing interest in enhancing PID-type controllers rather than abandoning them in favor of more complex designs.

To extend PID-based strategies to environments characterized by time-varying dynamics or limited model knowledge, early adaptive PID and auto-tuning methodologies were introduced. 
The seminal contributions of H\"agglund and \AA{}str\"om proposed relay-based auto-tuning and frequency–response identification for online gain adjustment~\cite{hagglundIndustrialAdaptiveControllers1991a}, later complemented by supervisory schemes designed to maintain robustness and prevent destabilizing interactions during adaptation~\cite{hagglundSupervisionAdaptiveControl2000a}. 
While these techniques have proven valuable, they remain largely heuristic and do not provide rigorous guarantees for nonlinear systems or in the presence of faults.

To address these limitations, Lyapunov-based adaptive PID and PID-like controllers were subsequently developed, providing stronger theoretical guarantees while retaining a structure compatible with practical implementation. 
The classical work of Chang~\cite{changSelftuningPIDControl2002} introduced direct Lyapunov-driven gain update laws to ensure closed-loop stability, and surveys~\cite{alexandrovAdaptivePIDControllers2014} emphasized the importance of combining nonlinear stability analysis with the architectural simplicity of PID control. 
More recent contributions include nonlinear adaptive PID schemes employing trimmable error transformations~\cite{xiangNonlinearAdaptivePID2025}, adaptive controllers with independently tunable gains~\cite{geAdaptivePIDControl2020}, and robust adaptive PD/PID-type methods for uncertain nonlinear dynamics~\cite{liRobustAdaptivePD2019}. 
Several nonlinear control paradigms have also incorporated PID-like structures in their implementation. This included:   backstepping controllers  ~\cite{kouraniPracticalGuidelineDesigning2022}; compensated PID-like controllers for non-affine plants~\cite{parkNovelCompensatedPID2025};  switching or sliding-mode augmentations ~\cite{kimPIDSwitchingAction2021}. The latter aimed at improving robustness while preserving a PID-compatible form.
In parallel, aspects of stability and boundedness in simple adaptive architectures were analyzed~\cite{barkanaGainConditionsConvergence2005}, highlighting the importance of preventing adaptive gains from growing unbounded.

Fault-tolerant extensions of these widely used architectures have also received considerable attention. 
Robust adaptive PID/PID-like schemes for MIMO systems with unknown control direction were reported~\cite{songRobustAdaptiveFaultTolerant2017}, while PID-like fault-tolerant controllers with time-varying scaling functions were proposed for actuator-fault scenarios~\cite{yangAdaptivePIDLikeFaultTolerant2024}. 
These and related works illustrate that PID-like designs—although often derived from more sophisticated nonlinear or adaptive principles—retain a practical structure that supports straightforward implementation in real-world systems.

Despite these advances, several limitations remain. Most adaptive PID and PID-like schemes rely on \emph{continuously active} adaptation laws, which can cause unnecessary gain drift, noise amplification, and loss of robustness under weak excitation. Only a small subset of existing approaches explicitly addresses process or actuator faults within a simple PID-like architecture while retaining Lyapunov-based guarantees. In addition, mechanisms designed to ensure bounded adaptive gains—such as gating and related regularization strategies—are largely absent from the current literature~\cite{hagglundSupervisionAdaptiveControl2000a,alexandrovAdaptivePIDControllers2014}.

The control architecture proposed in this work addresses these limitations by introducing a nonlinear adaptive design that preserves a PD-like structure while allowing the proportional gain and its coupled derivative component to adjust in response to fault-induced deviations\hl{, through a gated Lyapunov-based adaptation law that ensures boundedness while limiting unnecessary gain adjustments during nominal operation}.
More specifically, the article contributes by:
\begin{itemize}
    \item deriving a Lyapunov-based adaptive controller with a PD-like implementation featuring an adaptive proportional gain together with its damping-preserving derivative coupling;
    \item proposing a residual-based gating mechanism that activates adaptation only when the tracking error exceeds a prescribed threshold;
    \item incorporating a leakage term to guarantee bounded adaptive dynamics and establishing uniform global ultimate boundedness with explicit steady-state error bounds;
    \item demonstrating enhanced fault-rejection capabilities through a representative simulation study.
\end{itemize}

\section{Systems Class and Problem Formulation}

The following class of nonlinear systems in affine control form is considered:
\begin{equation}
    \ddot{x}
    = \xi(\mathbf{x})
    + g(\mathbf{x})\,u
    + \delta + f,
    \label{eq:system_affine}
\end{equation}
where $\mathbf{x} \triangleq [\,x\;\; \dot{x}\,]^\top\in\mathbb{R}^2$ is the state vector and $u\in\mathbb{R}$ is the control input. The vector field $g(\cdot)$ is assumed to be known whereas partial knowledge of $\xi(\cdot)$ may be available. \hl{The terms $\delta(t)$ and $f(t)$ denote unknown, time-varying, and bounded signals representing external disturbances and fault-induced perturbations, respectively.} More specifically, $\delta$ captures external perturbations during nominal operation, whereas $f$ describes abnormal or fault-induced deviations.

The formulation \eqref{eq:system_affine} encompasses a broad range of systems with intrinsic dynamics of this form, including mass--spring--damper mechanisms, pendulum-type systems, robotic joint actuators, unmanned aerial vehicle (UAV) altitude or translational dynamics, ground and marine vehicle motion models, and electrical RLC networks.  
This general representation serves as the basis for the fault-tolerant control scheme developed in the subsequent sections.

\begin{assumption}
\label{ass:system}
The system \eqref{eq:system_affine} satisfies the following conditions:
\begin{enumerate}
    \item The vector fields $\xi(\mathbf{x})$ and $g(\mathbf{x})$ are locally Lipschitz in $\mathbf{x}$ for all $\mathbf{x}\in\mathbb{R}^2$.

    \item The input gain $g(\mathbf{x})$ is uniformly bounded away from zero for all $\mathbf{x}\in\mathbb{R}^2$, i.e.,
    \[
        \lvert g(\mathbf{x}) \rvert \ge g_{\min} > 0 .
    \]

    \item There exist positive constant scalars $\bar{\delta}>0$ and $\bar{f}>0$ such that $\forall t \geq 0$ the disturbance and fault signals satisfy
    \[
        \left \vert \delta(t) \right \vert \le \bar{\delta}, \qquad \left \vert f(t) \right \vert \le \bar{f} .
    \]

    \item \hl{A nonnegative residual signal $r=r(t)\ge 0$ is available such that there exist constants $r_{\mathrm{th}}>0$ and $\varepsilon>0$ satisfying}
    \[
    \begin{aligned}
    r(t) &\ge r_{\mathrm{th}} \quad &&\text{if } \lvert f(t)\rvert > \varepsilon,\\
    r(t) &<  r_{\mathrm{th}} \quad &&\text{if } f(t) = 0 .
    \end{aligned}
    \]
    This property ensures that the fault component $f$ is isolable from the disturbance $\delta$ in a residual-based sense \cite{blankeDiagnosisFaultTolerantControl2016}.
    
\end{enumerate}
\end{assumption}

A twice continuously differentiable reference $x_r$ is considered, and the tracking error is defined as
\[
    e \triangleq x - x_r.
\]

A compensating term $\hat{\xi}(\mathbf{x})$ is used, representing the partial knowledge of the drift dynamics $\xi$, and the feedback–linearizing control input is selected as
\begin{equation}
    u = \frac{-\hat{\xi}(\mathbf{x}) - K_p e - K_d \dot{e} + \ddot{x}_r}{g(\mathbf{x})},
    \label{eq:FL_PD_short}
\end{equation}
with $K_p>0$ and $K_d>0$.  
This control law corresponds to a PD structure augmented with a compensation term that enforces a linearized closed-loop behavior. Substituting \eqref{eq:FL_PD_short} into \eqref{eq:system_affine} yields the closed-loop error dynamics
\begin{equation}
    \ddot{e} = -K_p e - K_d \dot{e} - d - f.
    \label{eq:err_short}
\end{equation}
The term
\[
    d \triangleq -\delta - \bigl(\xi(\mathbf{x}) - \hat{\xi}(\mathbf{x})\bigr)
\]
collects all non-fault uncertainty contributions (disturbances, modeling inaccuracies, and linearization or reconstruction errors).

Although the disturbance $d$ and the fault signal $f$ enter the dynamics in the same channel as additive signals and share identical mathematical properties (unknown, bounded, and unmatched to the PD loop), they play distinct roles in the control design.  
The disturbance $d$ persists during both nominal and faulty operation, whereas the fault component $f$ is detectable through the residual signal $r$, following standard fault diagnosis design procedures~\cite{blankeDiagnosisFaultTolerantControl2016}.  
Accordingly, the nominal PD controller is designed to counteract the influence of $d$, while the adaptive mechanism introduced later is responsible for mitigating the additional degradation caused by the fault signal $f$ whenever detected.

\begin{mdframed}[
    linewidth=0.5pt,
    linecolor=black,
    backgroundcolor=white,
    leftmargin=0pt,
    rightmargin=0pt,
    innerleftmargin=7pt,
    innerrightmargin=7pt,
    innertopmargin=5pt,
    innerbottommargin=5pt,
    skipabove=5pt,
    skipbelow=1pt
]
\hl{Therefore, the control problem addressed in this work is to design an adaptive control law that guarantees uniform ultimate boundedness of the tracking error in the presence of fault-induced deviations. Specifically, for any prescribed admissible bound $e_{\max}>0$, the controller parameters are to be designed such that there exists a finite $T>0$ for which the tracking error satisfies}
\[
\lvert e(t) \rvert \le e_{\max}, \qquad \forall\, t \ge T .
\]
\end{mdframed}

\section{Proposed Solution}
\label{sec:proposed_solution}

The central idea of the proposed fault--tolerant controller is to augment the nominal PD law with an adaptive gain mechanism capable of increasing the proportional (and correspondingly the differential) gains whenever the tracking error grows beyond nominal expectations.  
The adaptation mechanism is constructed with the form
\begin{equation}
    \label{eq:adapt_idea}
    \dot{\Delta K_p}
    = d(s_p)\,\Psi(\mathbf{e})- \sigma\,\Delta K_p,
\end{equation} 
where $\Delta K_p$ is the deviation from the nominal proportional gain, $\mathbf{e} \triangleq [\,e\;\; \dot{e}\,]^\top$ is the error state vector, $d(s_p)$ is a \emph{gating function} that modulates the controller gain adaptation rate based on an appropriately selected time-varying signal $s_p$, $\sigma$ is a leakage coefficient~\cite{slotineAppliedNonlinearControl1991a}, and $\Psi(\mathbf{e})$ is an adequate function of the error which will be defined later. Ideally, $\Psi(\mathbf{e})$ should be strictly positive for all $(e,\dot e)\neq (0,0)$, ensuring that the sign of $\dot{\Delta K_p}$ is dictated
exclusively by $d(s_p)$.  
The latter serves two purposes: it activates the gain update only under faulty conditions, and it scales the update rate according to the severity and persistence of the
deviation, ensuring the correct responsiveness under fault. The adequate coefficients discharge during recovery is instead guaranteed by the $\sigma$ coefficient, as demonstrated later.

\subsection*{Gating Mechanism}

To effectively implement this modulation, the gating function $d(s_p)$ is
constructed considering a nonnegative auxiliary signal $s_p$, obtained by integrating a measure of the excess of the tracking error beyond a prescribed activation threshold. Two levels of tolerance are introduced: an activation
threshold $e_{\mathrm{th}}>0$ and a performance limit $e_{\mathrm{lim}}>0$
satisfying
\[
    0 < e_{\mathrm{th}} \lesssim e_{\mathrm{lim}}.
\]
The threshold $e_{\mathrm{th}}$ determines when the adaptive mechanism becomes
active, whereas $e_{\mathrm{lim}}$ denotes the maximum admissible tracking
error. This separation ensures that adaptation is triggered sufficiently early
to maintain $|e| \le e_{\mathrm{lim}}$ throughout fault onset and recovery.

The activation signal is defined as
\[
    \xi_p \triangleq \vert e \vert - e_{\mathrm{th}},
\]
which quantifies how much the tracking error exceeds the activation
threshold. This quantity is integrated over time to form the gating variable
$s_p$, which accumulates the effect of persistent deviations. Accounting for the
residual signal $r$ (Assumption~\ref{ass:system}) and enforcing non-negativity
through projection, the evolution of $s_p$ is given by
\begin{equation}
    \dot{s}_p =
    \begin{cases}
        \max\{\,r\,\xi_p,\,0\,\}, & s_p = 0,\\[2mm]
        r\,\xi_p, & s_p > 0,
    \end{cases}
    \label{eq:sp_projection_clean}
\end{equation}
so that $s_p$ increases when fault-induced deviations occur ($r>0$ and
$\xi_p>0$), decreases only when $s_p>0$ and $\xi_p<0$, and is prevented from
becoming negative.

The gating variable modulates the adaptive gain through the strictly positive scaling function
\begin{equation}
    d(s_p) \triangleq \rho +
    \begin{cases}
        \eta_{p}\, s_p, & \xi_p \ge 0,\\[2mm]
        \eta_{n}\, s_p, & \xi_p < 0,
    \end{cases}
    \label{eq:dsp_clean}
\end{equation}
with design parameters $\rho>0$, $\eta_p>0$, and $\eta_n>0$.  
The structure in \eqref{eq:dsp_clean} ensures that $d(s_p)$ grows proportionally to $s_p$ when the error exceeds $e_{\mathrm{th}}$ and decreases (with a possibly different slope) when the error falls below this threshold, while always remaining strictly positive.  
As a result, the adaptation gain evolves rapidly during degradation of performance and more slowly during recovery, which is instrumental in preserving the desired performance level.

\begin{remark}
The function $d(s_p)$ satisfies $d(s_p)\ge\rho>0$ for all $s_p\ge 0$, is continuous and piecewise continuously differentiable, and is non-decreasing whenever $|e|\ge e_{\mathrm{th}}$.  
\end{remark}

\section{Main Result}
\label{sec:main_result}

Following the gain–modulation principle introduced in
Section~\ref{sec:proposed_solution}, this section details the complete
closed-loop controller.

The nominal PD action is augmented with adaptive gain
increments $\Delta K_p$ and $\Delta K_d$, yielding the modified control input
\begin{equation}
    u = \frac{-\hat{\xi}(\mathbf{x})
              - (K_p + \Delta K_p)e
              - (K_d + \Delta K_d)\dot{e}
              + \ddot{x}_r}{g(\mathbf{x})},
    \label{eq:main_control_law_revised}
\end{equation}
where $K_p>0$ and $K_d>0$ denote the baseline gains.

To preserve the desired closed-loop damping characteristics while maintaining a simple and robust adaptive structure, the derivative gain increment is linearly coupled to the proportional increment according to
\begin{equation}
    \Delta K_d = c_{\mathrm{lin}}\,\Delta K_p,
    \label{eq:Kd_coupling_revised}
\end{equation}
with a constant $c_{\mathrm{lin}}>0$. The motivation for this linear coupling and the derivation of $c_{\mathrm{lin}}$ are provided in Appendix~\ref{app:gain_coupling}.

The proportional gain increment evolves according to the adaptation law
\begin{equation}
    \dot{\Delta K_p}
    = d(s_p)\,\Psi(e,\dot{e})
      - \sigma\,\Delta K_p,
    \label{eq:adaptive_law_main}
\end{equation}
where $d(s_p)$ is the gating-dependent scaling function defined in \eqref{eq:dsp_clean}, and $\sigma$ the leakage term later introduced to ensure stability. This update rule coincides with the generic gain–modulation structure introduced in
Section~\ref{sec:proposed_solution}.

The function $\Psi(e,\dot{e})$ is given explicitly by
\begin{equation}
    \Psi(e,\dot{e})
    \triangleq p_{12} e^2 + (p_{22} + c_{\mathrm{lin}}p_{12})\,e\,\dot{e}
       + c_{\mathrm{lin}}p_{22} (\dot{e})^2,
    \label{eq:Psi_definition}
\end{equation}
where $p_{ij}$ denotes the entries of the matrix $P$, the unique symmetric
positive-definite solution of the Lyapunov equation
\begin{equation}
    A^{\top} P + P A = -Q,
    \qquad
    Q=\mathrm{diag}(q_{11},q_{22})> 0,
    \label{eq:Lyap_equation}
\end{equation}
related to the nominal closed-loop error dynamics in~\eqref{eq:err_short}. The closed-loop system matrix $A$ is given by
\begin{equation}
    \dot{\mathbf{e}} = A\mathbf{e}-B_f\,(d+f), \quad A = \begin{bmatrix}
    0 & 1\\
    -K_p & -K_d
\end{bmatrix}, \; B_f = \begin{bmatrix}
    0\\1
\end{bmatrix}.
\end{equation}

The update law \eqref{eq:adaptive_law_main} increases the gain whenever
fault-induced deviations dominate (corresponding to $d(s_p)$ growing through the
gating mechanism) and decreases the gain at a controlled rate when the tracking
error re-enters the activation region.  
This ensures that the adaptive increments $\Delta K_p$ and $\Delta K_d$ respond
accordingly to the severity and persistence of the performance degradation.

Substituting the control law
\eqref{eq:main_control_law_revised}--\eqref{eq:Kd_coupling_revised}
into the system dynamics yields the closed-loop error model
\begin{equation}
    \ddot{e}
    = -(K_p + \Delta K_p)\,e
      -(K_d + c_{lin}\Delta K_p)\,\dot{e}
      - d - f,
    \label{eq:main_error_dynamics_revised}
\end{equation}
which represents a second-order system subject to the disturbance $d$ and the
fault signal $f$.

\begin{theorem}
\label{thm:main_revised}
Consider the system \eqref{eq:system_affine} under
Assumptions~\ref{ass:system}, equipped with the gating mechanism
\eqref{eq:sp_projection_clean}--\eqref{eq:dsp_clean} and the adaptive controller
\eqref{eq:main_control_law_revised}--\eqref{eq:adaptive_law_main} with the gain
coupling \eqref{eq:Kd_coupling_revised}.  
Then all closed-loop signals in \eqref{eq:main_error_dynamics_revised} remain
uniformly bounded, and the tracking error $e$ is uniformly ultimately bounded.
In particular, for any prescribed performance limit $e_{\mathrm{lim}}>0$, there
exist design parameters $\rho$, $\eta_p$, $\eta_n$, $\sigma$,
$c_{\mathrm{lin}}$ and a positive scalar $T > 0$ such that the corresponding ultimate bound satisfies
\begin{equation}
    |e(t)| \le e_{\mathrm{lim}}, \; \forall t \geq T
    \label{eq:ft_cond}
\end{equation}
even in the presence of any disturbance $d$ and
fault signal $f$ consistent with Assumption~\ref{ass:system}.  Furthermore, the
adaptive gain increments $\Delta K_p$ and $\Delta K_d$ increase sufficiently
during fault-induced deviations to ensure that the regulated error remains
within the admissible limit $e_{\mathrm{lim}}$.
\end{theorem}

\begin{proof}
Select $P$ as in \eqref{eq:Lyap_equation}, and consider the Lyapunov candidate
\begin{equation}
    V(\mathbf{e},\Delta K_p)
    = \tfrac{1}{2}\,\mathbf{e}^\top P \mathbf{e}
      + \frac{(\Delta K_p)^2}{2\,d(s_p)},
    \label{eq:Lyap_candidate_proof}
\end{equation}
which is positive definite in $(\mathbf{e},\Delta K_p)$ since $P > 0$ and
$d(s_p) \ge \rho > 0$ by construction.

The time derivative of $V$ along the trajectories of the closed-loop system is
\begin{align}
    \dot{V}
    = &-\tfrac{1}{2}\,\mathbf{e}^\top Q \mathbf{e}
      - \Delta K_p\,\Psi(e,\dot{e})
      + (d+f)\,\Phi(e,\dot{e})\\
      &+ \frac{\Delta K_p}{d(s_p)}\,\dot{\Delta K_p}
      - \frac{(\Delta K_p)^2}{2\,d^2(s_p)}\,\dot{d}(s_p),
    \label{eq:Vdot_before_adaptation_correct}
\end{align}
where $\Psi(e,\dot{e})$ is given by \eqref{eq:Psi_definition} and the function $\Phi(e,\dot{e}) \triangleq -\mathbf{e}^\top P B$. The matrix $B = [\,0\,\;1\,]^\top$ represents the input channel of the
linearized open-loop system and is defined by:
\begin{equation}
    \dot{\mathbf{e}}
    =
    \begin{bmatrix}
        0 & 1\\
        0 & 0
    \end{bmatrix}\mathbf{e}
    \;+\;
    B\,\big( \ddot{x} - \ddot{x}_r \big).
\end{equation}

Next, substitute the adaptation law \eqref{eq:adaptive_law_main}
into \eqref{eq:Vdot_before_adaptation_correct}.  
The terms involving $\Delta K_p \Psi(e,\dot{e})$ cancel by construction, and one
obtains
\begin{equation}
\dot V
= -\tfrac{1}{2}\mathbf{e}^\top Q\mathbf{e}
  - \frac{(\Delta K_p)^2}{d(s_p)}
    \!\left(\sigma+\tfrac{\dot d(t)}{2\,d(s_p)}\right)
  + (d+f)\,\Phi(e,\dot{e}).
    \label{eq:Vdot_collected_DK}
\end{equation}

The term associated with $(\Delta K_p)^2$ in \eqref{eq:Vdot_collected_DK} may
become positive whenever $\dot{d}(s_p) < 0$.  
To ensure net energy dissipation in the adaptive channel, the leakage gain
$\sigma$ is selected such that
\begin{equation}
\label{eq:sigma_condition_final}
\sigma \;>\; \frac{|\dot d(t)|}{2\,d(s_p)}
\;>\; \frac{\eta_n\,r(t)\,e_{\max}}{2\,\rho},
\end{equation}
where \(d_{\min}=\rho\) is the minimum value of \(d(s_p)\) and
\(\eta_n\) is the slope active when \(|e|<e_{\mathrm{th}}\).

The remaining term $(d+f)\,\Phi(e,\dot{e})$ can be upper-bounded using
Young’s inequality:
\begin{equation}
    \big| (d+f)\,\Phi(\mathbf{e}) \big|
\le\; \frac{q}{4}\,\|\mathbf{e}\|^2 \;+\; \frac{1}{q}\,\|P B_f\|^2\,\big(|d|+|f|\big)^2,
\label{eq:young_bound}
\end{equation}
where $q \triangleq \lambda_{\min}(Q)$. As a result,
\begin{align}
\dot V
&\le -\tfrac{1}{2}\,q\,\|\mathbf{e}\|^2 + \frac{q}{4}\,\|\mathbf{e}\|^2
     + \frac{1}{q}\,\|P B_f\|^2\,\big(|d|+|f|\big)^2 \nonumber\\
&= -\tfrac{q}{4}\,\|\mathbf{e}\|^2
   + \frac{\|P B_f\|^2}{q}\,\big(|d|+|f|\big)^2 .
\label{eq:Vdot_ISS}
\end{align}

Since $|d(t)|+|f(t)|$ is bounded, this inequality 
implies that the set
\[
S =\Bigl\{\mathbf e\in\mathbb{R}^2\,\Big|\ \|\mathbf e\|
\le
\frac{2}{q}\,\|P B_f\|\,
\big(\|d\|_\infty+\|f\|_\infty\big)
\Bigr\}
\]
is globally attractive since, outside this set, the Lyapunov derivative is strictly negative and drives the trajectories back inside. Consequently,
\begin{equation}
\label{eq:limsup_bound}
\limsup_{t\to\infty}\|\mathbf e(t)\|
\;\le\;
\frac{2}{q}\,
\|P B_f\|\,
\big(\|d\|_\infty+\|f\|_\infty\big).
\end{equation}

Since $|e(t)| \le \|\mathbf e(t)\|$, evaluating $\|P B_f\|$ from the solution
of \eqref{eq:Lyap_equation} and replacing
$p_{12}$ and $p_{22}$ with their expressions in terms of $q_{11}$
and $q_{22}$—the diagonal entries of $Q$—yields
\begin{equation}
\begin{aligned}
|e(t)|
&\le \frac{\|d\|_\infty+\|f\|_\infty}
{q\,(K_p + \Delta K_p)} \\
&\quad \times
\sqrt{
q_{11}^2 +
\left[
\frac{q_{11} + q_{22}(K_p + \Delta K_p)}
{K_d + c_{\mathrm{lin}}\Delta K_p}
\right]^2
}
\end{aligned}
\label{eq:bound_cond}
\end{equation}
This expression---whose detailed derivation is provided in Appendix~\ref{app:ultimate_bound}---makes explicit how the ultimate tracking error
depends on the gain increment $\Delta K_p$ for given bounds on the disturbance
and fault signals. By imposing $|e(t)| < e_{\mathrm{lim}}$, one may solve
\eqref{eq:bound_cond} implicitly for $\Delta K_p$ to obtain the minimum gain
increase required to guarantee the fault-tolerant condition \eqref{eq:ft_cond} to be satisfied.
\end{proof}

\section{Parameter Influence and Tuning Guidelines}
\label{sec:tuning}

The nominal performance of the closed loop is determined by selecting the gains
$K_p$ and $K_d$ based on conventional steady-state and transient requirements.
The adaptive scheme contribution related to the adaptive term $\Delta K_p$---and the consequent $\Delta K_d$ to maintain the plant's original damping characteristic---requires the tuning of the parameters appearing in the adaptation
law, namely $\eta_p$, $\eta_n$, $\rho$, and $\sigma$, as well as the coefficients
of $P$ through the choice of $Q$. \hl{The following analysis provides both practical tuning guidelines and
a detailed discussion of the influence of each design parameter on closed-loop performance.}

\subsection{Selection of the Matrix \texorpdfstring{$Q$}{Q}}
\label{sec:tuning_Q}

The matrix $Q > 0$ influences the adaptive behavior through the solution
$P$ of the Lyapunov equation.  
In particular, the entries $p_{12}$ and $p_{22}$ determine the quadratic expression $\Psi(e,\dot e)$, which can be written compactly as
\begin{equation}
    \Psi(e,\dot e)
    = \mathbf{e}^\top
    \underbrace{
    \begin{bmatrix}
        p_{12} & \tfrac{1}{2}(p_{22}+c_{\mathrm{lin}}p_{12}) \\[4pt]
        \tfrac{1}{2}(p_{22}+c_{\mathrm{lin}}p_{12}) & c_{\mathrm{lin}}p_{22}
    \end{bmatrix}
    }_{=: M}
    \mathbf{e}.
\end{equation}

Ideally, it should satify the condition
\[
    \Psi(e,\dot e)\ge 0
    \qquad \forall\,(e,\dot e)\in\mathbb{R}^2,
\]
so that the quantity $d(s_p)$ uniquely determines the sign of the adaptive
increment through $\xi_p$.  
This is achieved when $M > 0$, which—since $p_{12}>0$—reduces to the
condition $\det M \geq 0$.  
It can be shown with straightforward calculations that $\det M\leq 0$. Hence, the only choice is $\det M = 0$ which yields 
\begin{equation}
\label{eq:q_cond}
    p_{22} = c_{\mathrm{lin}} p_{12}
    \quad\Longleftrightarrow\quad
    q_{22} = 
    \frac{(c_{\mathrm{lin}}K_d^{\mathrm{eff}} - 1)}{K_p^{\mathrm{eff}}}\, q_{11},
\end{equation}
using the relation between $P$ and $Q$ implied by \eqref{eq:Lyap_equation}, with
$K_p^{\mathrm{eff}} = K_p + \Delta K_p$ and
$K_d^{\mathrm{eff}} = K_d + c_{\mathrm{lin}}\Delta K_p$.

When the closed-loop parameters satisfy $c_{\mathrm{lin}}K_d^{\mathrm{eff}} > 1$,
this choice guarantees
\[
    \Psi(e,\dot e) = p_{12}(e + c_{\mathrm{lin}}\dot e)^2 \;\ge\; 0.
\]

If instead $c_{\mathrm{lin}}K_d^{\mathrm{eff}} < 1$, the condition above implies
a negative value of $q_{22}$, which is not admissible since $Q > 0$.  
In this case, global non negativity of $\Psi$ cannot be enforced.  A suitable
design guideline is therefore to select
\[
    q_{11} \gg q_{22},
\]
which results in the $\Psi$ function being positive over the vast majority of the
$(e,\dot e)$-plane.  
This reduces the region in which $\Psi$ becomes negative and typically yields
more favourable adaptive behavior\hl{, highlighting the influence of the matrix selection on the regularity of the gain adaptation}
(see Appendix~\ref{app:Q_sel}).

\subsection{Positive Modulation Tuning: $\boldsymbol{\eta_p}$ and $\boldsymbol{\rho}$}
\label{sec:tuning_ds_p}

The coefficient $\eta_p>0$ determines the rate at which the adaptive gain
$\Delta K_p$ increases when the tracking error exceeds the activation threshold. Thus, its selection is critical, as it governs the speed of fault compensation and the
ability of the controller to achieve the desired effective proportional gain in fault conditions. An indicative value for the amplitude of the nominal perturbation and the expected fault can help in deriving an indicative value for $\eta_p$, based on a simplified average of the adaptation law, thereby providing practical tuning guidelines and easing commissioning.

Let $d_{\mathrm{nom}}$ denote the nominal disturbance amplitude and let $\Delta d_f^{\%}$ be the increase in the perturbation amplitude. The proportional gain increase $\Delta K_p^\star$ which is required to maintain the error within the band under faulty conditions can be estimated from the commonly used approximate steady–state requirement\cite{astromPIDControllersTheory1995}
\[
    \Delta K_p^\star
    = \frac{(1 + \Delta d_f^{\%})\,d_{\mathrm{nom}}}{e_{\mathrm{lim}}} - K_p.
\]
Assume now that the desired value $\Delta K_p^\star$ must be reached within a
transition time $T_{\Delta}$ after the fault starts. Integrating the adaptation law \eqref{eq:adaptive_law_main} over that period leads to
\begin{equation}
    \begin{aligned}
        \Delta K_p^\star
= \int_0^{T_\Delta} \!\!\left(\, \rho + \eta_p\,\Psi\,s_p\right)\,\mathrm{d}\tau
> \frac{\alpha_\Psi\,\alpha_e}{2}\eta_p\,\overline{\Psi}\,\overline{\xi}_p\,T_\Delta^{2},
    \end{aligned}
\end{equation}
where the following approximations have been utilized:
\begin{itemize}
    \item $\overline{\Psi} > \alpha_\Psi\,p_{12}\,e_{\text{lim}}^{2}$ to only depend on known values, with the coefficient $\alpha_\Psi \in [1.5-3]$ to partially compensate the presence of the other terms in $\Psi$ and avoid over-conservatism,
    \item $s_p \approx \overline{\xi}_pT_\Delta > \alpha_e\,e_{\text{lim}}T_\Delta$ assuming that the error will overshoot---on average---an additional conservative $\alpha_e \in [0.1- 0.2]$  portion of $e_{\text{lim}}$ during the transient 
    \item $\rho > 0$ as it is only needed to ensure $d(s_p)>0$ and a correct definition of $V$.
\end{itemize}
Substituting these approximations gives the analytical relation
\begin{equation}
\label{eq:eta_p_formula}
    \eta_p
    \;\approx\;
    \frac{2\,\Delta K_p^\star}{\alpha_{\Psi}\,\alpha_e\,p_{12}\,e_{\mathrm{lim}}^3\,T_{\Delta}^2}.
\end{equation}
\hl{This relation explicitly links the choice of $\eta_p$ to the desired adaptation speed and fault-tolerant performance, providing} a reliable initial tuning guideline. The selection of the $\alpha_e$ and $\alpha_{\Psi}$ are related to the nature of the system and to the expected performance while under fault.

The tuning of $\rho$ reduces to selecting a sufficiently small value, namely
$\rho \ll \eta_p$, while avoiding values too close to zero.  
Choosing $\rho$ excessively small would force the leakage gain
$\sigma$ to satisfy $\sigma \approx \infty$ in order to meet the stability
condition, which is neither practical nor desirable.

\subsection{Negative Modulation Tuning: $\boldsymbol{\eta_n}$ and $\boldsymbol{\sigma}$}
\label{sec:tuning_ds_n}

A separate coefficient $\eta_n$ is introduced for the gain discharge phase
because the stability condition imposes $\sigma > \sigma_{\min} \propto \eta_n$.
If $\eta_n$ were comparable to $\eta_p$, the resulting $\sigma_{\min}$ would be
large, producing an undesirably rapid decay of the adaptive gain and, more
critically, significantly limiting the positive growth of $\Delta K_p$ during
fault compensation.

By selecting $\eta_n \ll \eta_p$ instead, the discharge dynamics naturally become slower,
and the minimum value of $\sigma$ required to ensure stability is reduced.
Should a faster discharge be desired, $\sigma$ can be increased starting from the
value obtained from \eqref{eq:sigma_condition_final} until satisfactory decay
performance is achieved. \hl{The selection of these parameters therefore determines the trade-off between rapid gain recovery and effective fault compensation.}

\section{Case Study: Positioning System}

\begin{figure}[tp]
    \centering
    \includegraphics[width=\columnwidth]{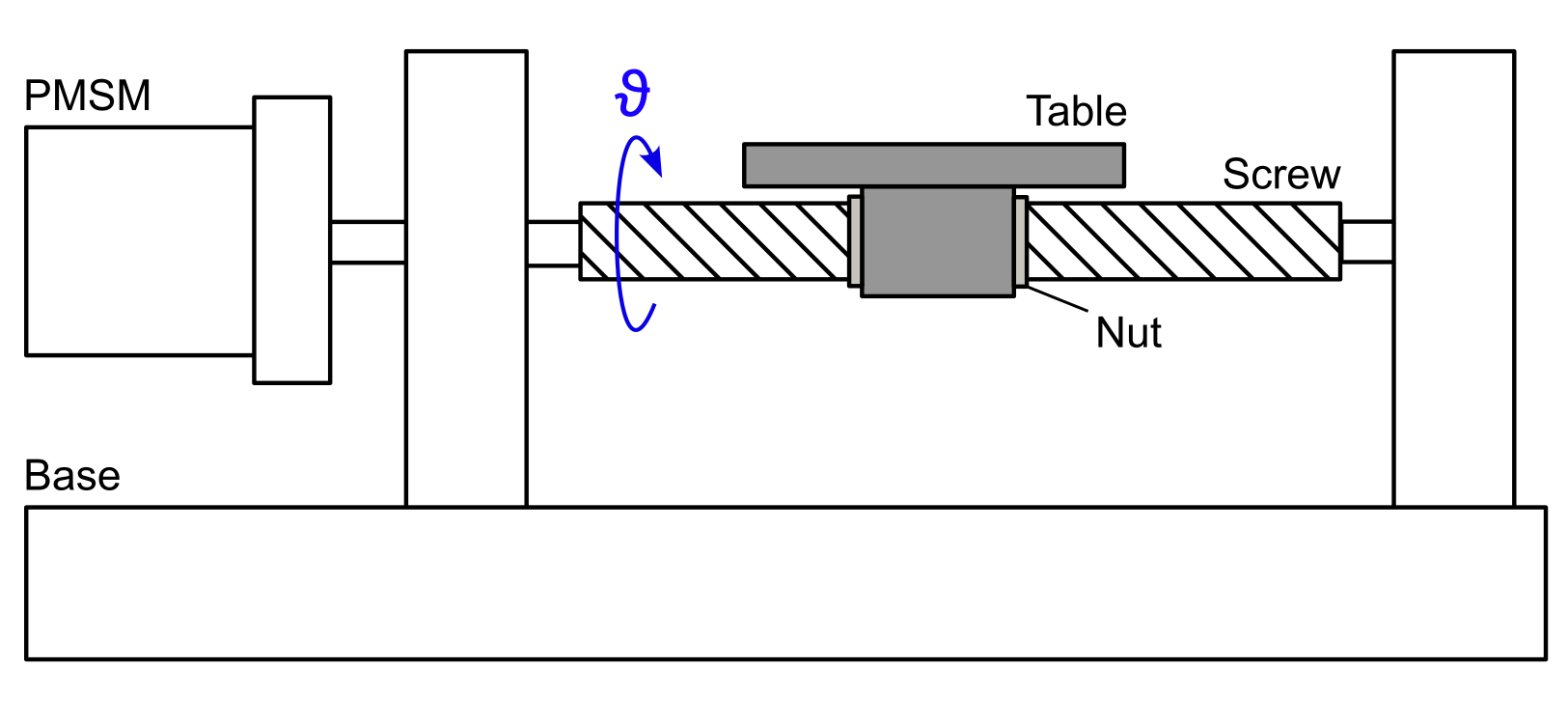}
    \caption{Precision positioning system: schematics.}
    \label{fig:case_study}
\end{figure}

The effectiveness of the proposed adaptive fault-tolerant controller is assessed in simulation for a precision positioning system consisting of an angular position regulation
loop for a permanent-magnet synchronous motor (PMSM), whose schematics is presented in Figure~\ref{fig:case_study}.

\begin{table}[bp]
\centering
\footnotesize
\caption{System, baseline, and adaptive control parameters.}
\label{tab:dimpa_params}
\begin{tabular}{p{0.35\columnwidth}cc}
\toprule
\textbf{Parameter} & \textbf{Symbol} & \textbf{Value} \\
\midrule
\multicolumn{3}{c}{\textit{System parameters}} \\[3pt]
Moment of inertia & $J$ & $8.31\times10^{-4}$ kg$\cdot$m$^{2}$ \\
Transition sharpness & $\alpha$ & $1\times10^{5}$ \\
Coulomb friction coeff. & $T_{\mathrm{cm}}$ & $0.0223$ \\
Viscous friction coeff. & $\beta_m$ & $0.0016$ \\[3pt]

\midrule
\multicolumn{3}{c}{\textit{Baseline control}} \\[3pt]
Proportional gain & $K_p$ & $450$ \\
Derivative gain & $K_d$ & $35$ \\
Lyapunov matrix & $Q$ & $\begin{bmatrix}1&0\\0&2.84\times10^{-4}\end{bmatrix}$ \\[3pt]

\midrule
\multicolumn{3}{c}{\textit{Adaptive control}} \\[3pt]
Adaptation threshold & $e_{\max}$ & $0.08$ \\
Error limit & $e_{\mathrm{lim}}$ & $0.10$ \\
Gating offset & $\rho_{s_p}$ & $30$ \\
Positive slope coefficient & $\eta_{s_p}^+$ & $2.2\times10^{8}$ \\
Negative slope coefficient & $\eta_{s_p}^-$ & $20$ \\
Leakage term & $\sigma$ & $2.93\times10^{-2}$ \\
$K_p$ limit (secant approx.) & $\Delta K_{p,\max}$ & $K_{p}$\\
Gain  ratio & $c_{lin}$ & $3.22\times10^{-2}$ \\
[3pt]

\midrule
\multicolumn{3}{c}{\textit{Tuning Rules}} \\[3pt]
Nominal disturbance bound & $d_{\mathrm{nom}}$ & $T_{\mathrm{cm}}$ \\
Expected fault & $\Delta d_f^{\%}$ & 1.5 \\
Adaptation time & $T_\Delta$ & 4 s \\
Regressor weight & $\alpha_{\Psi}$ & 2 \\
Integral term weight & $\alpha_{e}$ & 0.1 \\
\bottomrule
\end{tabular}
\end{table}

The electrical dynamics governed by the inner current control loop is neglected
due to the significantly faster response compared with the mechanical
dynamics, which can be expressed as
\begin{equation}
    \ddot{\theta}
    = \frac{1}{J}\left(u - F_{\mathrm{fr}} - f\right),
\end{equation}
where $\theta$ denotes the angular position and $J$ the total inertia.  
The friction torque $F_{\mathrm{fr}}$ is modeled as the sum of Coulomb and
viscous contributions,
\begin{equation}
    F_{\mathrm{fr}}
    = F_c\,\tanh(\alpha\,\dot{\theta}) + B_{v}\,\dot{\theta},
\end{equation}
with $\alpha$ determining the steepness of the smooth sign-function
approximation.  
The fault $f$ represents the effect of debris entering the mechanical
transmission, producing an increase in the static friction coefficient \hl{$\Delta F_c$} of
approximately 100--150\%. \hl{The corresponding fault term is defined as}
\begin{equation}
    f = \Delta F_c \tanh(\alpha\,\dot{\theta}).
\end{equation}
This scenario constitutes a representative benchmark for fault-tolerant control,
as friction is difficult to model precisely and highly challenging to estimate.

Let $\theta_r$ be a time-varying reference to be tracked by the angular position
$\theta$, and let $e_\theta=\theta-\theta_r$ denote the associated tracking
error.  
The baseline feedback-linearizing controller is given by
\begin{equation}
    u = -J(K_p\,e_\theta + K_d\, \dot e_\theta).
\end{equation}
Defining the lumped disturbance term $d = F_{\mathrm{fr}}$, the nominal
closed-loop system coincides with \eqref{eq:err_short}, with $K_p$ and $K_d$
tuned to satisfy the steady-state error requirements under nominal operating
conditions.

The system and baseline control parameters, together with the additional tuning
parameters related to the adaptation law, are summarized in
Table~\ref{tab:dimpa_params}, following the configuration reported
in previous work~\cite{papageorgiouHighAccuracyNonlinear2017}.  
The values of $Q$ and $\eta_p$ are selected according to the proposed tuning
guidelines, with the $c_{lin}$ coefficient being calculated via the secant approximation (see Appendix~\ref{app:gain_coupling}).
 
\subsection{Simulation Results}

\begin{figure}
    \centering
    \includegraphics[width=\columnwidth]{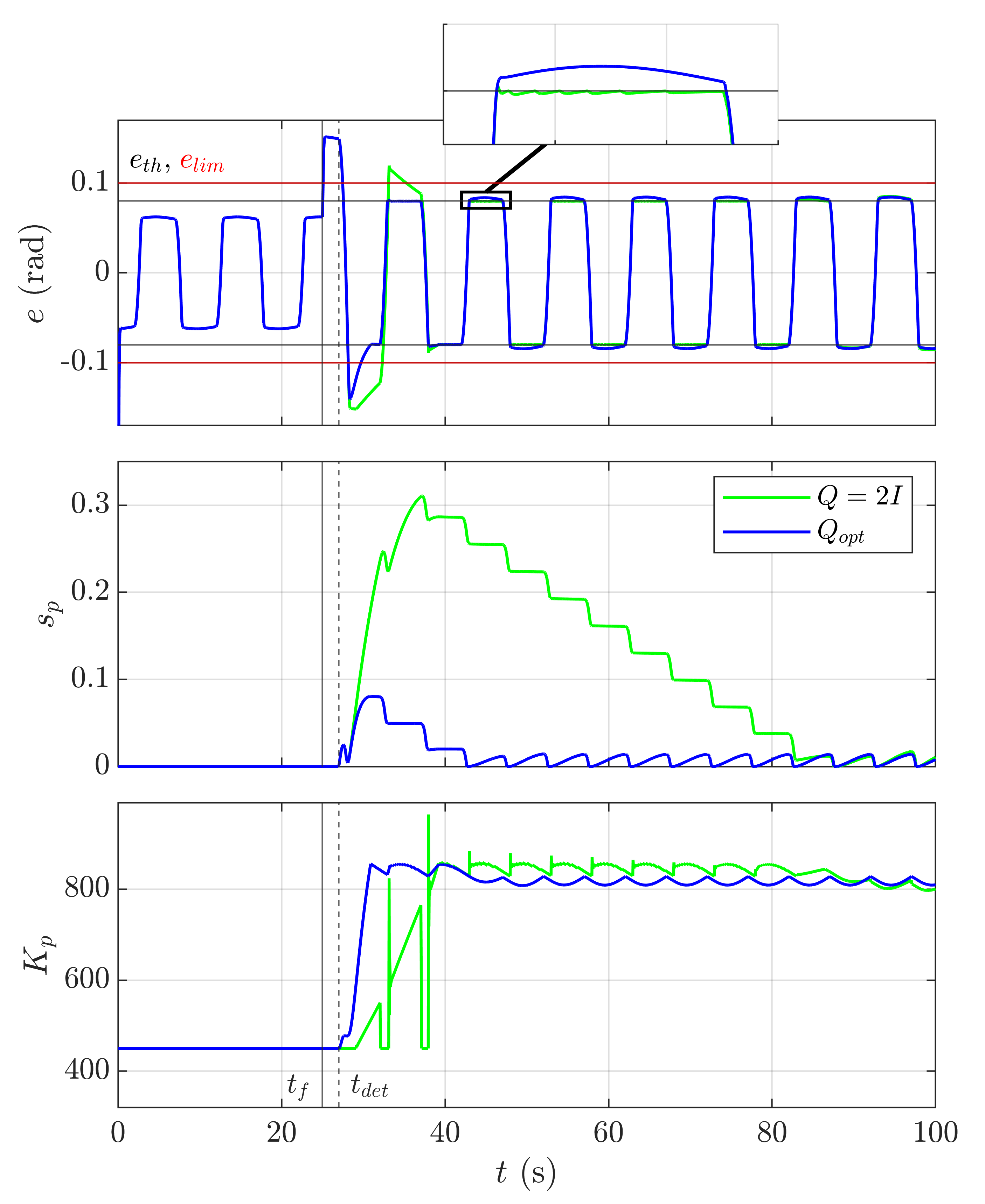}
    \caption{Gain increase during fault-tolerant operation, for different Q values.}
    \label{fig:increase}
\end{figure}

The closed-loop response obtained for the faulty scenario introduced earlier in this section is shown in Fig.~\ref{fig:increase}. Two design choices for the matrix $Q$ are compared: the optimal selection---computed according to the condition in~\eqref{eq:q_cond}---and the classical choice $Q = 2I$, both capable of driving the error back within the prescribed precision limits. The latter yields a less predictable evolution of the adaptive gain, caused by sign changes of the $\psi$ function that induce sudden decreases in the gain even when $\xi_p > 0$. In contrast, the optimal selection ensures that such function remains strictly positive, resulting in a smooth and rapid gain increase whenever the tracking error exceeds the threshold $e_{th}$.

Although both designs lead to comparable steady-state behavior, the non-optimal choice of $Q$ produces a markedly longer transient. This occurs because $s_p$ accumulates a larger contribution during intervals in which the $\psi$ function drives the gains downward. As a consequence, the adaptive gain attains higher values in the latter part of the transient, keeping the tracking error closer to the threshold during this phase, as evident from the zoomed view.

For the optimal design, the transient duration is approximately $4\,\mathrm{s}$, which closely matches the value prescribed during the tuning procedure. This agreement confirms that the proposed tuning guidelines provide an effective and reliable indication for selecting the adaptation parameters.

To evaluate the gain–discharge phase and to illustrate the influence of the remaining parameters, the scenario in which the fault is removed after some time—representing, for instance, debris suddenly exiting the bearing—is considered. In this case, both $\rho$ and $\eta_n$ are chosen to be several orders of magnitude smaller than $\eta_p$, as discussed in Section~\ref{sec:tuning}. As previously noted, selecting $\rho$ too small—or, equivalently, choosing $\eta_n$ excessively large—forces the minimum permissible value of $\sigma$ to increase, which in turn degrades the performance of the gain–increase phase.

\begin{figure}
    \centering
    \includegraphics[width=\columnwidth]{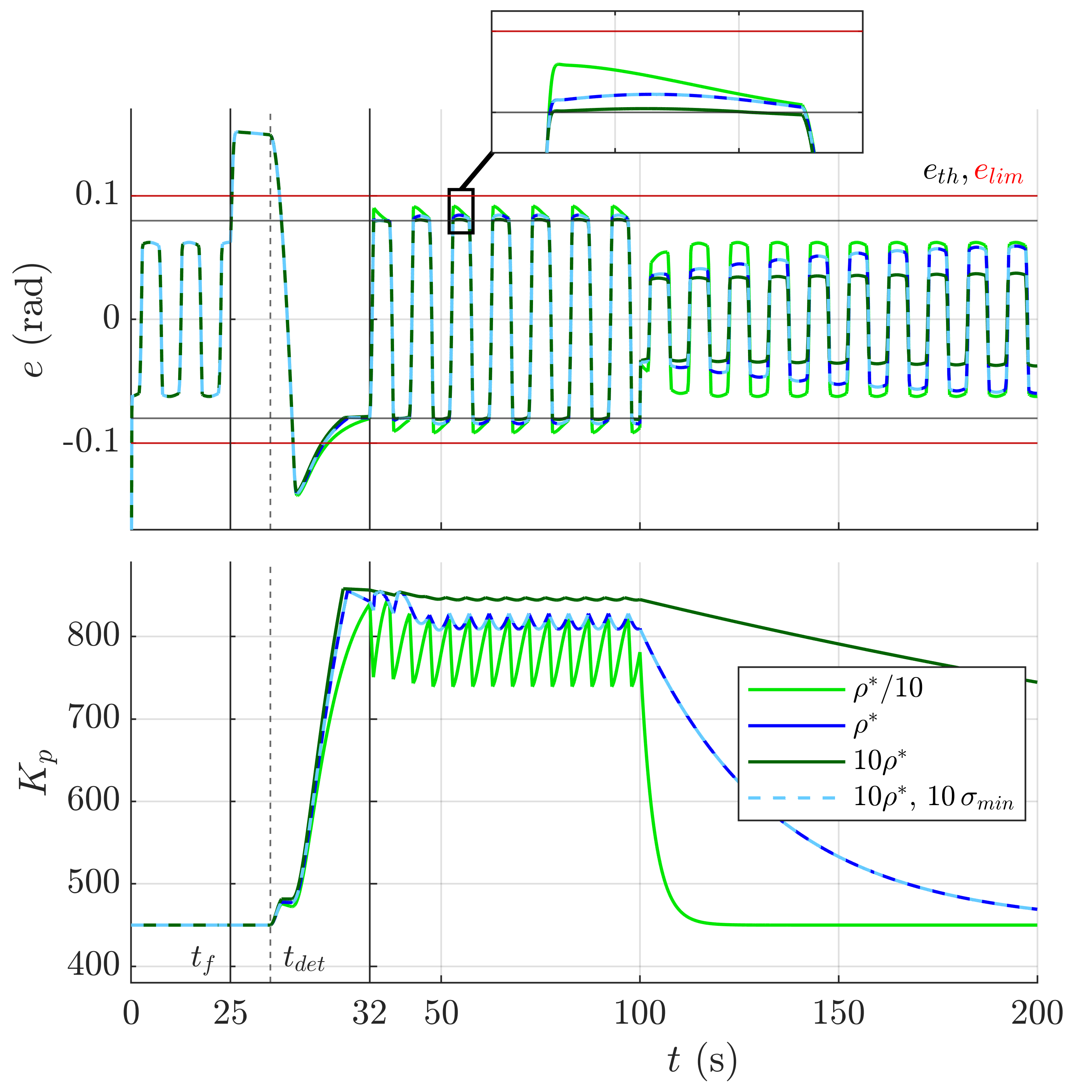}
    \caption{Gain discharge with different tuning variations.}
    \label{fig:discharge}
\end{figure}

Figure~\ref{fig:discharge} shows the behavior of the algorithm during the gain–decrease period for different values of $\rho$. The same considerations apply to $\eta_n$, but in the opposite direction. Unless otherwise specified, $\sigma$ is taken to be only slightly larger than the minimum value required by the stability condition. The value $\rho^\ast$, reported in Table~\ref{tab:dimpa_params}, represents the final tuning choice, providing a suitable compromise between achieving a sufficiently fast gain discharge after the fault disappears and limiting the oscillation of the tracking error around the threshold.

Reducing $\rho$ below this threshold increases the minimum admissible value of $\sigma$, thereby degrading performance; in such cases, a reduction of $\eta_n$ can partially mitigate this effect. Conversely, selecting $\rho$ significantly larger weakens the discharge action: although this choice yields excellent performance during fault-tolerant operation, it results in higher steady-state gains and limits the controller’s ability to return to its nominal tuning once the system reverts to normal conditions. In this scenario, $\sigma$ can be safely increased until the discharge dynamics match those of the optimal configuration.

In summary, the tuning of these coefficients is not unique, and comparable performance can be obtained from different combinations of $\eta_n$, $\sigma$, and $\rho$. Depending on the application—specifically, the likelihood of the fault disappearing—the designer may prioritize either a tighter, smoother fault‐tolerant response or a more balanced behavior that preserves adequate discharge capability.

\section{Conclusions}

This work presented a residual–based adaptive gain–modulation strategy for a class of
nonlinear systems in affine control form.  
The proposed approach augments a nominal feedback–linearizing PD controller with
an adaptive mechanism activated through a residual–driven gating variable.  
By integrating the error excess above a prescribed threshold and modulating the
adaptation through a piecewise linear metric, the method selectively increases
the control gains only when fault–induced performance degradation is detected,
while remaining inactive during nominal operation.  
This construction provides a transparent and implementable fault–tolerant
enhancement to standard PD control laws.

The correct gated adaptation law is derived from a Lyapunov–based stability analysis, which also enabled the formulation of explicit tuning indications for the gating parameters, the leakage term, and the choice of the $Q$ matrix. The resulting guidelines yield a coherent and predictable tuning procedure,
clarifying the trade–offs between adaptation speed, discharge behavior, and the regularity of the $\Psi$ function.

The effectiveness of the proposed controller was demonstrated on a precision
positioning system subject to sudden variations in friction.  
The adaptive gain successfully counteracted large fault–induced deviations,
keeping the tracking error within the prescribed performance threshold.  
The benefit of selecting an appropriate anisotropic $Q$ matrix---ensuring a
predominantly positive $\Psi$ function---was confirmed to yield shorter and more regular
transients.

Future work includes extending the approach to PI–type structures—for instance,
velocity control through an integral error—and to PID–type designs suitable for
higher–order systems.  
Another promising direction is the integration of the gated adaptation mechanism
with disturbance–estimation schemes, made natural by the delayed activation of
the adaptive component, which leaves room for faster, non–conflicting corrective
actions within the inner performance band.

\appendix
\section{Derivative-Gain Coupling}
\label{app:gain_coupling}

Consider the nominal PD–closed loop
\begin{equation}
    \ddot e + K_d \dot e + K_p e = -\,d - f ,
    \label{eq:app_nomCL}
\end{equation}
whose natural frequency and damping ratio satisfy
\begin{equation}
    \omega_0 = \sqrt{K_p}, 
    \qquad 
    \xi = \frac{K_d}{2\sqrt{K_p}}.
    \label{eq:app_xi_def}
\end{equation}
Thus, increasing $K_p$ without modifying $K_d$ reduces the closed-loop damping.  
Preserving the damping ratio $\xi$ exactly would require enforcing the nonlinear coupling
\begin{equation}
    \phi(\Delta K_p)
    \triangleq
    K_{d,0}\!\left(
        \sqrt{1+\frac{\Delta K_p}{K_{p,0}}} - 1
    \right),
    \label{eq:phi_def}
\end{equation}
which follows directly from the definition of $\xi$ in \eqref{eq:app_xi_def}.
A linear approximation of \eqref{eq:phi_def} is adopted to preserve the damping ratio with good accuracy over the relevant adaptation range, while yielding a constant coupling coefficient. This choice avoids introducing unnecessary nonlinearities in the adaptive law and facilitates the systematic definition of a positive regressor $\Psi$. The linear coefficient $c_{\mathrm{lin}}$ obtained by approximating
$\phi(\Delta K_p)$ in one of the following ways:
\begin{itemize}
    \item \textbf{Tangent approximation (at $\Delta K_p=0$):}
    \[
        c_{\mathrm{tan}}
        = \left.\frac{\partial \phi}{\partial \Delta K_p}\right|_{\Delta K_p=0}
        = \frac{K_{d,0}}{2K_{p,0}}.
    \]
    This yields a damping increase and is appropriate when the nominal
    loop is underdamped.
    \vspace{0.3cm}
    \item \textbf{Secant approximation (over $[0,\Delta K_{p,\max}]$):}
    \begin{equation}
        c_{\mathrm{sec}}
        =
        \frac{\phi(\Delta K_{p,\max})}{\Delta K_{p,\max}}
        \label{eq:csec_general}
    \end{equation}
    which better distributes the approximation error over the expected
    gain–increase interval and is appropriate for mildly overdamped
    nominal behavior.
\end{itemize}

\section{Ultimate Bound Derivation}
\label{app:ultimate_bound}

Let $P=P^\top>0$ be the solution of the Lyapunov equation
\begin{equation}
    A^\top P + P A = -Q, 
    \qquad Q=\mathrm{diag}(q_{11},q_{22})>0,
    \label{eq:Lyap_appendix}
\end{equation}
associated with the nominal closed-loop matrix
\[
    A =
    \begin{bmatrix}
        0 & 1 \\
        -(K_p+\Delta K_p) & -(K_d+c_{\mathrm{lin}}\Delta K_p)
    \end{bmatrix},
    \qquad
    B_f=
    \begin{bmatrix}
        0\\1
    \end{bmatrix}.
\]

Solving the Lyapunov equation yields
\[
    p_{12}
    =
    \frac{q_{11}}{2(K_p+\Delta K_p)},
    \qquad
    p_{22}
    =
    \frac{q_{11}+q_{22}(K_p+\Delta K_p)}
         {2(K_d+c_{\mathrm{lin}}\Delta K_p)\,(K_p+\Delta K_p)} .
\]

Therefore
\[
    \|P B_f\|^2
    =
    p_{12}^2 + p_{22}^2
    =
    \frac{1}{4}
    \left[
        \frac{q_{11}^2}{(K_p+\Delta K_p)^2}
        +
        \frac{
            \bigl(q_{22} + q_{11}/(K_p+\Delta K_p)\bigr)^2
        }{
            (K_d + c_{\mathrm{lin}}\Delta K_p)^2
        }
    \right].
\]

Substituting $\|P B_f\|$ into the steady-state bound
and using $|e|\le\|e\|$ gives the expression quoted in
equation~\eqref{eq:bound_cond}.

\section{Q-Matrix Selection and Positivity of the Regressor}
\label{app:Q_sel}

\begin{figure}[bp]
    \centering
    \includegraphics[width=\columnwidth]{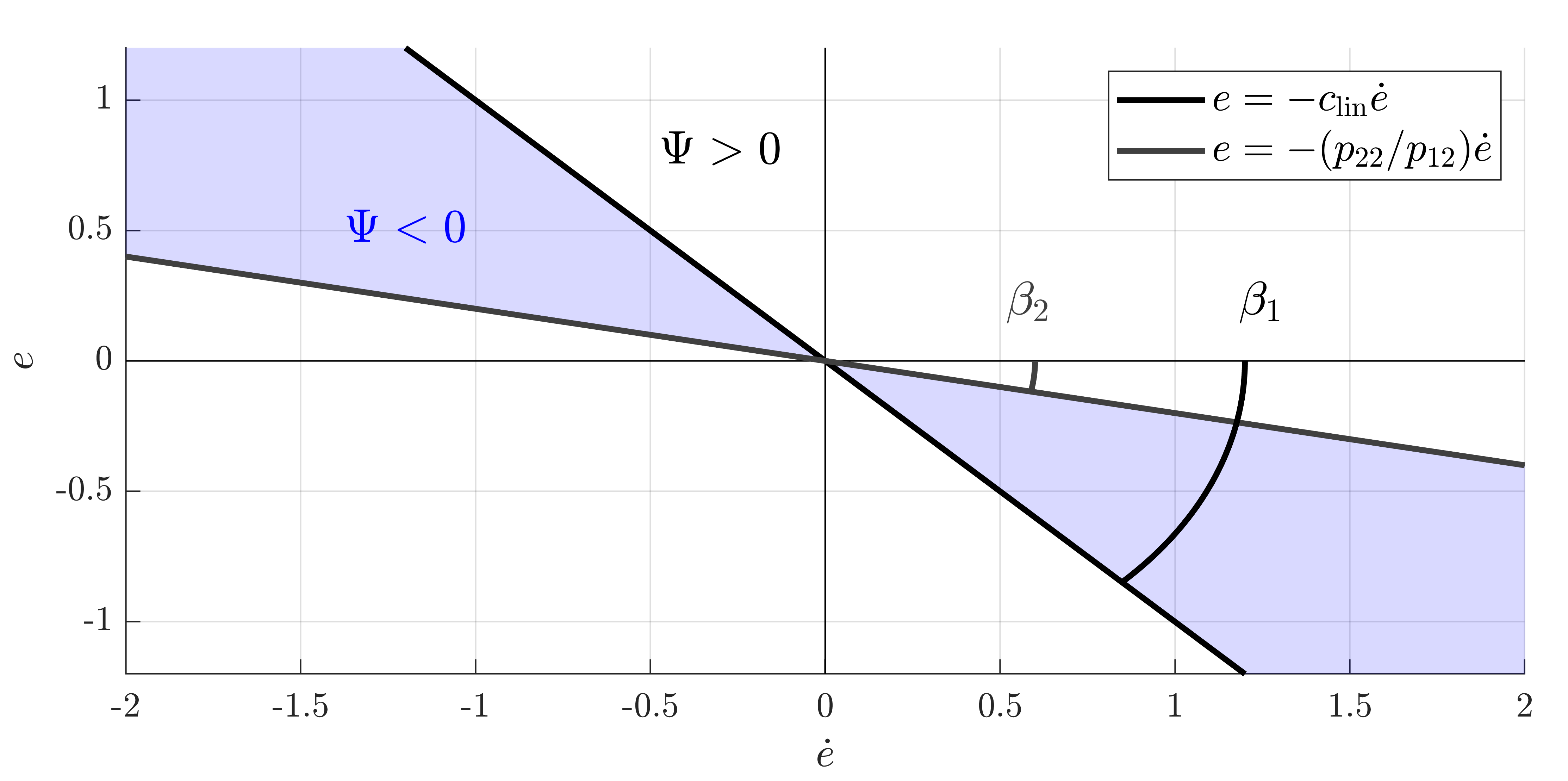}
    \caption{$\Psi(e,\dot e)$ cones in the $(e,\dot e)$-plane.}
    \label{fig:e_e_dot_plane}
\end{figure}

If $c_{\mathrm{lin}}K_{d}^{\mathrm{eff}}<1$, the optimal expression implies
$q_{22}<0$, which is inadmissible since $Q>0$.
In this regime, global positivity of $\Psi$ is impossible.  The goal is
therefore to minimize the region in which $\Psi<0$. Using the expressions for $p_{12}$ and $p_{22}$,
$\Psi(e,\dot e)$ factorizes exactly as
\[
    \Psi(e,\dot e)
    = p_{12}
      \bigl(e + c_{\mathrm{lin}}\dot e\bigr)
      \bigl(e + \tfrac{p_{22}}{p_{12}}\dot e\bigr),
\]
so that $\Psi=0$ on the two lines
\[
    \ell_{1}: e=-c_{\mathrm{lin}}\dot e,
    \qquad
    \ell_{2}: e=-\tfrac{p_{22}}{p_{12}}\dot e;
\]
where $\beta_1 \triangleq \arctan(c_{\mathrm{lin}})$ and
$\beta_2 \triangleq \arctan(p_{22}/p_{12})$ define their angles with respect to the $\dot e$–axis, as illustrated in Figure~\ref{fig:e_e_dot_plane}.
$\Psi$ is negative exactly in the conical region 
$$
    L = \left\lbrace (e,\dot{e})\in\mathbb{R}^2 \big | \bigl(e + c_{\mathrm{lin}}\dot e\bigr)
      \bigl(e + \tfrac{p_{22}}{p_{12}}\dot e\bigr)  < 0\right\rbrace.
$$
Since
\[
    \frac{p_{22}}{p_{12}}
    =
    \frac{1}{K_{d}^{\mathrm{eff}}}
    \left(1 + K_{p}^{\mathrm{eff}}\frac{q_{22}}{q_{11}}\right),
\]
increasing $q_{11}/q_{22}$ moves $\ell_{2}$ closer to $\ell_{1}$, thereby
shrinking the cones where $\Psi<0$. Thus, when
$c_{\mathrm{lin}}K_{d}^{\mathrm{eff}}<1$, the recommended choice is
\[
    q_{11} \gg q_{22},
\]
which ensures that $\Psi(e,\dot e)$ remains positive on almost the
entire $(e,\dot e)$--plane,
significantly improving the adaptive behavior by confining the
$\Psi<0$ regions to narrow cones.


\bmsection*{Conflict of interest}
The authors declare no potential conflict of interests.

\bibliography{es_bib}

@article{alexandrovAdaptivePIDControllers2014,
  title = {Adaptive {{PID}} Controllers: {{State}} of the Art and Development Prospects},
  shorttitle = {Adaptive {{PID}} Controllers},
  author = {Alexandrov, A. G. and Palenov, M. V.},
  year = 2014,
  month = feb,
  journal = {Automation and Remote Control},
  volume = {75},
  number = {2},
  pages = {188--199},
  issn = {0005-1179, 1608-3032},
  doi = {10.1134/S0005117914020027},
  urldate = {2025-11-24},
  copyright = {http://www.springer.com/tdm},
  langid = {english}
}

@book{astromPIDControllersTheory1995,
  title = {{{PID}} Controllers: Theory, Design, and Tuning},
  shorttitle = {{{PID}} Controllers},
  author = {{\AA}str{\"o}m, Karl J. and H{\"a}gglund, Tore},
  year = 1995,
  edition = {2. ed},
  publisher = {Instrument Society of America},
  address = {Research Triangle Park, NC},
  isbn = {978-1-55617-516-9},
  langid = {english}
}

@article{barkanaGainConditionsConvergence2005,
  title = {Gain Conditions and Convergence of Simple Adaptive Control},
  author = {Barkana, Itzhak},
  year = 2005,
  month = feb,
  journal = {International Journal of Adaptive Control and Signal Processing},
  volume = {19},
  number = {1},
  pages = {13--40},
  issn = {0890-6327, 1099-1115},
  doi = {10.1002/acs.830},
  urldate = {2025-11-18},
  copyright = {http://doi.wiley.com/10.1002/tdm\_license\_1.1},
  langid = {english}
}

@book{blankeDiagnosisFaultTolerantControl2016,
  title = {Diagnosis and {{Fault-Tolerant Control}}},
  author = {Blanke, Mogens and Kinnaert, Michel and Lunze, Jan and Staroswiecki, Marcel},
  year = 2016,
  publisher = {Springer},
  address = {Berlin, Heidelberg},
  doi = {10.1007/978-3-662-47943-8},
  urldate = {2023-03-22},
  isbn = {978-3-662-47942-1 978-3-662-47943-8},
  langid = {english}
}

@article{changSelftuningPIDControl2002,
  title = {A Self-Tuning {{PID}} Control for a Class of Nonlinear Systems Based on the {{Lyapunov}} Approach},
  author = {Chang, Wei-Der and Hwang, Rey-Chue and Hsieh, Jer-Guang},
  year = 2002,
  month = feb,
  journal = {Journal of Process Control},
  volume = {12},
  number = {2},
  pages = {233--242},
  issn = {0959-1524},
  doi = {10.1016/S0959-1524(01)00041-5},
  urldate = {2025-11-24}
}

@article{geAdaptivePIDControl2020,
  title = {Adaptive {{PID Control}} for {{Second Order Nonlinear Systems}}},
  author = {Ge, Zhui and Liu, Feng and Meng, Lingjun},
  year = 2020,
  month = aug,
  journal = {2020 Chinese Control And Decision Conference (CCDC)},
  pages = {2926--2931},
  issn = {1948-9447},
  doi = {10.1109/CCDC49329.2020.9164463},
  urldate = {2025-11-24}
}

@article{hagglundIndustrialAdaptiveControllers1991a,
  title = {Industrial Adaptive Controllers Based on Frequency Response Techniques},
  author = {H{\"a}gglund, Tore and Johan {\AA}str{\"o}m, Karl},
  year = 1991,
  month = jul,
  journal = {Automatica},
  volume = {27},
  number = {4},
  pages = {599--609},
  issn = {0005-1098},
  doi = {10.1016/0005-1098(91)90052-4},
  urldate = {2025-11-24}
}

@article{hagglundSupervisionAdaptiveControl2000a,
  title = {Supervision of Adaptive Control Algorithms},
  author = {H{\"a}gglund, Tore and {\AA}str{\"o}m, Karl Johan},
  year = 2000,
  month = aug,
  journal = {Automatica},
  volume = {36},
  number = {8},
  pages = {1171--1180},
  issn = {0005-1098},
  doi = {10.1016/S0005-1098(00)00026-1},
  urldate = {2025-11-24}
}

@article{kimPIDSwitchingAction2021,
  title = {{{PID}} with a {{Switching Action Controller}} for {{Nonlinear Systems}} of {{Second-order Controller Canonical Form}}},
  author = {Kim, MyoungHo and Lee, Sung-Uk},
  year = 2021,
  month = jul,
  journal = {International Journal of Control, Automation and Systems},
  volume = {19},
  number = {7},
  pages = {2343--2356},
  issn = {1598-6446, 2005-4092},
  doi = {10.1007/s12555-020-0346-4},
  urldate = {2025-11-24},
  langid = {english}
}

@article{kouraniPracticalGuidelineDesigning2022,
  title = {A {{Practical Guideline}} for {{Designing}} and {{Tuning Adaptive Backstepping Controllers}} for a {{Class}} of {{Second-Order Systems}} Based on {{PID Similarity}}},
  author = {Kourani, Ahmad and Daher, Naseem},
  year = 2022,
  month = dec,
  journal = {International Journal of Dynamics and Control},
  volume = {10},
  number = {6},
  pages = {1829--1846},
  issn = {2195-268X, 2195-2698},
  doi = {10.1007/s40435-022-00922-8},
  urldate = {2025-11-24},
  langid = {english}
}

@article{liRobustAdaptivePD2019,
  title = {Robust {{Adaptive PD Tracking Control}} of {{Nonlinear Uncertain Systems}} with {{Application}} to {{Unmanned Aerial Vehicles}}},
  author = {Li, Wenqiao and Tan, Shilei and Cao, Qian and Lai, Junfeng},
  year = 2019,
  month = may,
  journal = {2019 3rd International Symposium on Autonomous Systems (ISAS)},
  pages = {210--215},
  doi = {10.1109/ISASS.2019.8757697},
  urldate = {2025-11-24}
}

@phdthesis{papageorgiouHighAccuracyNonlinear2017,
  title = {High {{Accuracy Nonlinear Control}} and {{Estimation}} for {{Machine Tool Systems}}},
  author = {Papageorgiou, Dimitrios},
  year = 2017,
  address = {Department of Electrical Engineering},
  school = {Technical University of Denmark}
}

@article{parkNovelCompensatedPID2025,
  title = {A {{Novel Compensated PID Controller Ensuring Lyapunov Stability}} for {{Highly Uncertain}}, {{Nonaffine}}, and {{Time-Varying Nonlinear Systems}}},
  author = {Park, Jang-Hyun and Lee, Dong-Ho},
  year = 2025,
  journal = {IEEE Access},
  volume = {13},
  pages = {83897--83904},
  issn = {2169-3536},
  doi = {10.1109/ACCESS.2025.3568438},
  urldate = {2025-11-24},
  copyright = {https://creativecommons.org/licenses/by/4.0/legalcode}
}

@book{slotineAppliedNonlinearControl1991a,
  title = {Applied {{Nonlinear Control}}},
  author = {Slotine, Jean-Jacques E. and Li, Weiping},
  year = 1991,
  publisher = {Prentice-Hall},
  googlebooks = {HddxQgAACAAJ},
  isbn = {978-0-13-040049-9},
  langid = {english}
}

@article{songRobustAdaptiveFaultTolerant2017,
  title = {Robust {{Adaptive Fault-Tolerant PID Control}} of {{MIMO Nonlinear Systems With Unknown Control Direction}}},
  author = {Song, Yongduan and Huang, Xiucai and Wen, Changyun},
  year = 2017,
  month = jun,
  journal = {IEEE Transactions on Industrial Electronics},
  volume = {64},
  number = {6},
  pages = {4876--4884},
  issn = {1557-9948},
  doi = {10.1109/TIE.2017.2669891},
  urldate = {2025-11-24}
}

@article{xiangNonlinearAdaptivePID2025,
  title = {Nonlinear {{Adaptive PID Control}} for {{Nonlinear Systems}}},
  author = {Xiang, Kaili and Song, Yongduan and Ioannou, Petros},
  year = 2025,
  month = oct,
  journal = {IEEE Transactions on Automatic Control},
  volume = {70},
  number = {10},
  pages = {7000--7007},
  issn = {1558-2523},
  doi = {10.1109/TAC.2025.3567565},
  urldate = {2025-11-24}
}

@article{yangAdaptivePIDLikeFaultTolerant2024,
  title = {Adaptive {{PID-Like Fault-Tolerant Control}} for {{Uncertain H-Type Gantry Stage With Actuator Failure}}},
  author = {Yang, Yana and Jin, Bokun and Li, Junpeng and Hua, Changchun},
  year = 2024,
  month = nov,
  journal = {IEEE Transactions on Industrial Electronics},
  volume = {71},
  number = {11},
  pages = {14440--14448},
  issn = {1557-9948},
  doi = {10.1109/TIE.2024.3370984},
  urldate = {2025-11-24}
}

\end{document}